\documentclass[iop,apj,revtex4]{emulateapj}
\usepackage{graphicx}
\usepackage{amsmath,amssymb}

\begin{document}
\title{Mirror instability in the turbulent solar wind}
\author{Petr Hellinger\altaffilmark{1,2}}
\email{petr.hellinger@asu.cas.cz}
\author{Simone Landi\altaffilmark{3,4}}
\author{Lorenzo Matteini\altaffilmark{5}}
\author{Andrea Verdini\altaffilmark{3}}
\author{Luca Franci\altaffilmark{3,6}}

\altaffiltext{1}{Astronomical Institute, CAS,
Bocni II/1401,CZ-14100 Prague, Czech Republic}
\altaffiltext{2}{Institute of Atmospheric Physics, CAS,
Bocni II/1401, CZ-14100 Prague, Czech Republic}
\altaffiltext{3}{Dipartimento di Fisica e Astronomia, Universit\`a degli Studi di Firenze Largo E. Fermi 2, I-50125 Firenze, Italy}
\altaffiltext{4}{INAF -- Osservatorio Astrofisico di Arcetri, Largo E. Fermi 5, I-50125 Firenze, Italy}
\altaffiltext{5}{Department of Physics, Imperial College London, London SW7 2AZ, UK}
\altaffiltext{6}{INFN -- Sezione di Firenze, Via G. Sansone 1, I-50019 Sesto F.no (Firenze), Italy}

\begin{abstract}

The relationship between a decaying strong turbulence and the mirror instability in a slowly expanding plasma
is investigated using two-dimensional hybrid expanding box simulations.
We impose an initial ambient magnetic
field perpendicular to the simulation box, and we start with a spectrum of
large-scale, linearly-polarized, random-phase Alfv\'enic fluctuations which have energy equipartition between kinetic and magnetic
fluctuations and vanishing correlation between the two fields. A turbulent cascade rapidly develops, magnetic field fluctuations
 exhibit a Kolmogorov-like power-law spectrum at large scales and
a steeper spectrum at sub-ion scales. The imposed expansion (taking a strictly transverse 
ambient magnetic field) leads to generation of
an important perpendicular proton temperature anisotropy that
eventually drives the mirror instability.
This instability generates large-amplitude, nonpropagating, compressible, pressure-balanced magnetic
structures in a form of magnetic enhancements/humps that
  reduce the perpendicular temperature anisotropy. 
\end{abstract}
\pacs{?}
\maketitle

\section{Introduction}

In situ observations in the solar wind, in planetary magnetosheaths, in the heliosheaths
and in other weakly collisional, generally turbulent astrophysical plasmas show 
isolated or wave-trains of compressible pressure-balanced structures
\citep{wintal95,stka07,tsural11,enrial13}. Many of these structures
are thought to be generated by the mirror instability driven
by the perpendicular particle temperature anisotropy
\citep{vesa58,hase69,hell07}.
The mirror instability has peculiar features. It generates nonpropagating
modes (at least in a plasma without differential streaming) and near threshold 
the unstable modes appears on fluid scales, i.e., on large scales with respect to
the particle characteristic scales. On the other hand, the instability is resonant,
the resonant particles (with nearly zero parallel velocities with respect to the ambient magnetic
field) have a strong influence on the instability growth rate
\citep{soki93}.

The nonlinear properties of the mirror instability are not well understood.
\cite{kuznal07b, kuznal07a} proposed a
 nonlinear model for this instability
near threshold,  based on a reductive
perturbative expansion of the Vlasov-Maxwell equations.
This model extends the mirror dispersion relation by
including the dominant nonlinear coupling whose effect is to
reinforce the mirror instability.
In this approach both the linear and nonlinear properties are 
strongly sensitive to the details of the proton distribution function \citep{calial08}.
In the perturbative nonlinear model the particle distribution is, however,
fixed. For bi-Maxwellian particle distribution functions the nonlinear
model predicts formation of magnetic depressions/holes at the nonlinear
stage of the instability. On the other hand, direct numerical simulations
typically show generation of magnetic enhancements/humps \citep{calial08}.
This behavior is in agreement with expectations based on the energy
minimization argument in the simplified framework of
usual anisotropic magnetohydrodynamics \citep{passal06}.
\cite{hellal09} attempted to combine the reductive
perturbative expansion approach with the quasilinear approximation \citep{shsh64}.
This combined model leads to a fast deformation of the proton distribution
function that modifies the (sign of the) nonlinear term, and, consequently,
magnetic humps are generated in agreement with fully self-consistent simulations.
The quasilinear approximation is however questionable in the case of coherent
structures and, moreover, one expects particle trapping to be important at the nonlinear
level of the mirror instability \citep{pantal95,rincal15}.

In situ observations in the terrestrial magnetosheath show
that mirror magnetic humps are typically observed in the mirror-unstable
plasma whereas in the mirror-stable plasma magnetic holes are more probable
  \citep{soucal08,genoal09}.
The two-dimensional hybrid expanding box simulation of a homogeneous
plasma system (with the magnetic field in the simulation box, without turbulent fluctuations but
with an expansion that drives the perpendicular temperature anisotropy \cite[cf.,][]{mattal12,hell17a}) 
of \cite{traval07b} predicts that in a high-beta plasma the
mirror modes are dominant, and, as the expansion pushes the system to lower betas, the system
becomes dominated by the proton cyclotron waves whereas the mirror mode structure continuously
disappear. The mirror modes, however, survive relatively far the stable region where
they are (linearly) damped; in the unstable region the mirror modes have the form
of magnetic humps and on the way to the stable region they transform to
 magnetic holes similar to the observations in the terrestrial
magnetosheath  \citep{genoal09,genoal11}. The transition from humps to holes is not
understood but is in agreement with the energetic arguments.

The mirror instability is usually investigated in homogeneous or weakly inhomogeneous plasmas
\citep{hase69,hell08,hercal13}. Behavior of this instability is a strongly turbulent (and strongly inhomogeneous) plasma
such as the solar wind one is an open problem. As in the case of the oblique
fire hose instability \citep{hellal15} one expects that the mirror instability coexists with
plasma turbulence if it is fast enough to compete with the turbulent cascade. 
In this paper we investigate properties of the mirror instability in 2-D expanding
box simulation where we include important turbulent plasma motion. The paper is
organized as follows:
Section~\ref{code} describes the numerical code, Section~\ref{simul} presents the simulation results,
and in Section~\ref{discussion} we discuss the obtained results. 

\section{Hybrid expanding box model}
\label{code}
In this paper we  test the relationship between
proton kinetic instabilities and plasma turbulence in the solar wind using
a hybrid expanding box model that allows
to studying self-consistently physical processes at ion scales.
In the hybrid expanding box model
a constant solar wind radial velocity $v_{sw}$ is assumed.
The radial distance $R$
is then
$R = R_0 (1+ t/t_{e0})$
where $R_0$ is the initial position and $t_{e0}=R_0/v_{sw}$
 is the initial value of the characteristic expansion time $t_e=R/v_{sw}=t_{e0}(1+t/t_{e0})$.
Transverse scales (with respect to the radial direction)
 of a small portion of plasma, co-moving with the solar
wind velocity, increase
$\propto R$.
The expanding box uses these co-moving coordinates,
approximating the spherical coordinates by the Cartesian ones \citep{hetr05}.
The model uses the hybrid approximation where electrons are
considered as a massless, charge neutralizing fluid and
ions are described by a particle-in-cell model \citep{matt94}.
Here we use the two-dimensional (2-D) version of the code,
fields and moments are defined on a 2-D $x$--$y$ grid
$2048 \times 2048$; periodic boundary conditions are assumed.
The spatial resolution is
$\Delta x=\Delta y=  0.25 d_{p0}$ where $d_{p0}=v_{A0}/\Omega_{\mathrm{p}0}$ is
the initial proton inertial length ($v_{A0}$: the initial Alfv\'en velocity,
$\Omega_{\mathrm{p}0}$: the initial proton gyrofrequency).
There are $4096$ macroparticles per cell
for protons which are advanced
with a time step $\Delta t=0.05/\Omega_{\mathrm{p}0}$
while the magnetic field
is advanced with a smaller time step $\Delta t_B = \Delta t/10$.
The initial ambient magnetic field is directed along the $z$ direction,
perpendicular to the simulation plane (that includes the radial direction $\|y$),
$\boldsymbol{B}_{0}=(0,0,B_{0})$, and we impose a
continuous expansion in $x$ and $z$ directions with the initial expansion time
$t_{e0}=10^4 \Omega_{\mathrm{p}0}^{-1}$.

Due to the expansion with the strictly transverse magnetic field
the ambient density and the magnitude of the ambient magnetic field
decrease as $\langle n\rangle \propto  R^{-2} $ while $\langle B\rangle \propto R^{-1}$ (the proton
inertial length $d_{\mathrm{p}}$ increases $\propto R$,
the ratio between the transverse sizes and $d_{\mathrm{p}}$ remains constant;
the proton gyrofrequency $\Omega_\mathrm{p}$ decreases as $\propto R^{-1}$).
A small resistivity $\eta$ is used to avoid accumulation of
cascading energy at grid scales;
we set $\eta = 0.002 \mu_0 v_{A0}^2/\Omega_{\mathrm{p}0}$
($\mu_0$ being the magnetic permittivity of vacuum).
The simulation is initialized with an isotropic 2-D spectrum
of modes with random phases, linear Alfv\'en polarization ($\delta
\boldsymbol{B} \perp \boldsymbol{B}_0$), and vanishing correlation between
magnetic and velocity fluctuations. These modes are in the range
 $0.02 \le k d_{\mathrm{p}} \le 0.2$ and have a flat one-dimensional (1-D) (omnidirectional) power spectrum with rms fluctuations $=0.25 B_0$.
We set initially
the parallel proton beta  $\beta_{p\|} =  3$
and the system is characterized by a  perpendicular temperature anisotropy
 $T_{p\perp}/T_{p\|}=1.6$; for these parameters the plasma system
is already unstable with respect to the mirror instability, however,
the geometrical constraints and the presence of relatively strong fluctuations
inhibit the growth of mirror modes.   
Electrons are assumed to be isotropic and isothermal with $\beta_{e} =   1$
at $t=0$.

\section{Simulation results}
\label{simul}

\begin{figure}[htb]
\centerline{\includegraphics[width=8cm]{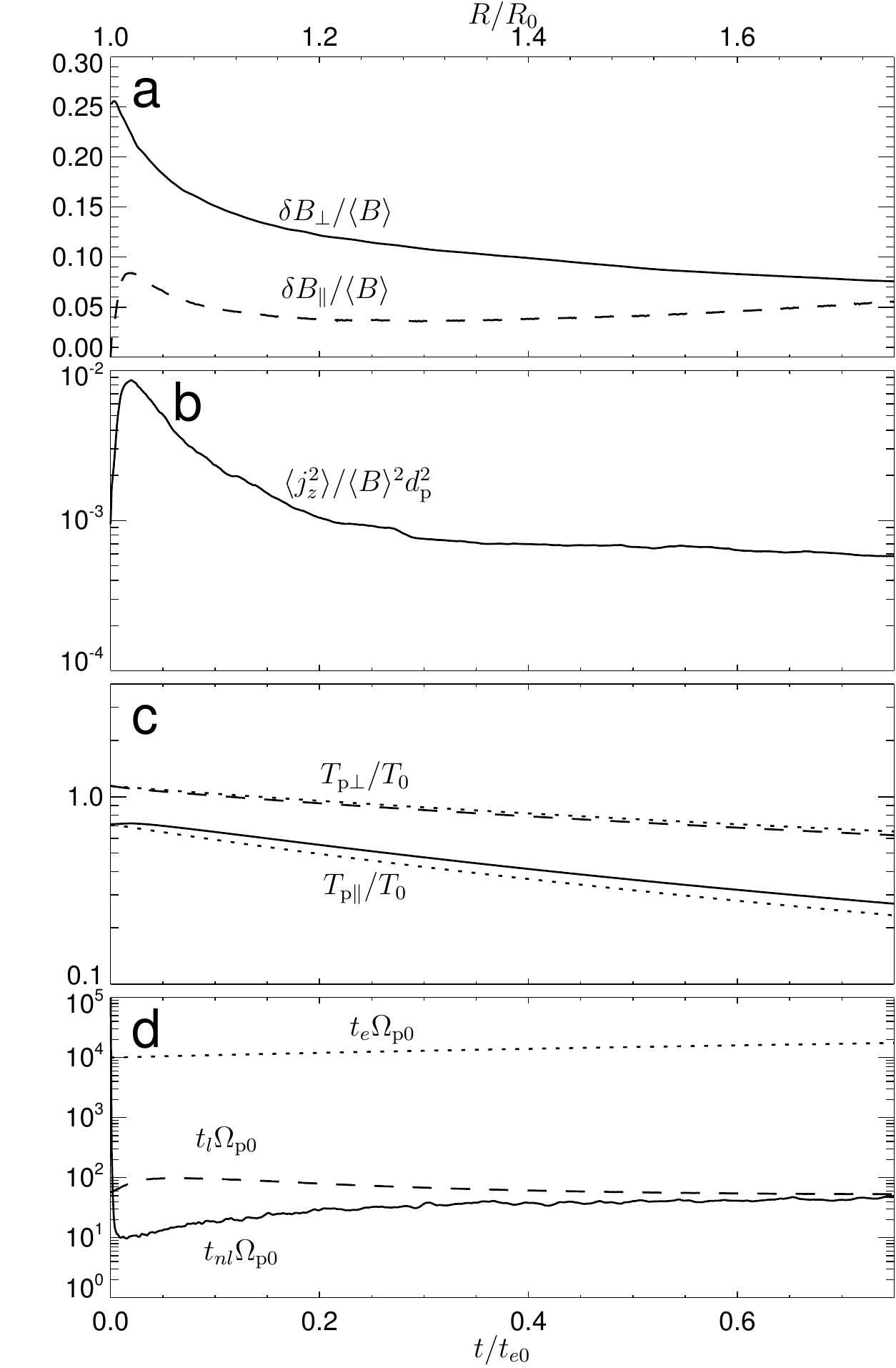}}
\caption{Time evolution of different quantities: (a) the fluctuating magnetic field
(solid) perpendicular $\delta B_\perp$ and (dashed)  parallel
 $\delta B_\|$ with respect to $\boldsymbol{B}_0$; (b)
the average squared parallel current $\langle j_z^2\rangle$; (c)
the parallel $T_{p\|}$ (solid line)
and perpendicular $T_{p\perp}$ (dashed line) proton temperatures
(the $\|$ and $\perp$ directions are here with respect to the local magnetic
field; the dotted lines denote the corresponding CGL predictions); (d)
(solid) the nonlinear eddy turnover time $t_{nl}$ at $k d_{\mathrm{p}}=1$
(dotted) the expansion time $t_e$, and (dashed) the linear time $t_l$ for the mirror
 instability.
\label{evol}}
\end{figure}

Figure~\ref{evol} shows the evolution of  different quantities in the
simulation as functions of time: the fluctuating magnetic field
(panel a, solid line) perpendicular $\delta B_\perp$ and (a, dashed)  parallel
 $\delta B_\|$ with respect to $\boldsymbol{B}_0$; (b)
the average squared parallel current $\langle j_z^2\rangle$;
the (c, solid) parallel $T_{p\|}$ 
and (c, dashed) perpendicular $T_{p\perp}$ proton temperatures
(the $\|$ and $\perp$ directions are here with respect to the local magnetic
field; the dotted lines on panel c denote the corresponding CGL predictions);
(d, solid) the nonlinear eddy turnover time $t_{nl}$ at $k d_{\mathrm{p}}=1$
(d, dotted) the expansion time $t_e$, and (d, dashed) the linear time $t_l=1/\gamma_{max}$ for the mirror
 instability; here $\gamma_{max}$ is the maximum growth rate of the mirror instability in the
corresponding homogeneous plasma with bi-Maxwellian protons.

Figure~\ref{evol} gives an overview of the simulation:
Initially, the parallel current fluctuations 
are generated, $\langle j_z^2\rangle$ (normalized to $\langle B\rangle^2/d_{\mathrm{p}}^2$) reaches a
maximum at $t\sim0.02 t_{e0}$ indicating the presence
of a well-developed turbulent cascade \citep{mipo09,serval15}.
After that the system is dominated by a decaying
turbulence, $\delta B_\perp/\langle B\rangle$ decreases. During the initial phase,
the compressible component $\delta B_\|$ is also generated, and then decays;
at later times, however, $\delta B_\| / \langle B\rangle$ stagnates and even increases.
This indicates generation of compressible fluctuations.
Figure~\ref{evol}c shows that the parallel and perpendicular temperatures roughly follow
the double adiabatic predictions. Figure~\ref{evol}d presents a comparison between characteristic
timescales; the longest timescale is the expansion time, the system is (on average) unstable
with respect to the mirror instability but the nonlinear eddy turnover time $t_{nl}$ at $k d_{\mathrm{p}}=1$
is faster than the mirror linear time but at later times the two timescales become comparable.
This may be favorable for generation of compressible mirror modes that may be responsible
for the increasing compressible magnetic fluctuations.

\begin{figure*}[htb]
\centerline{\includegraphics[width=14cm]{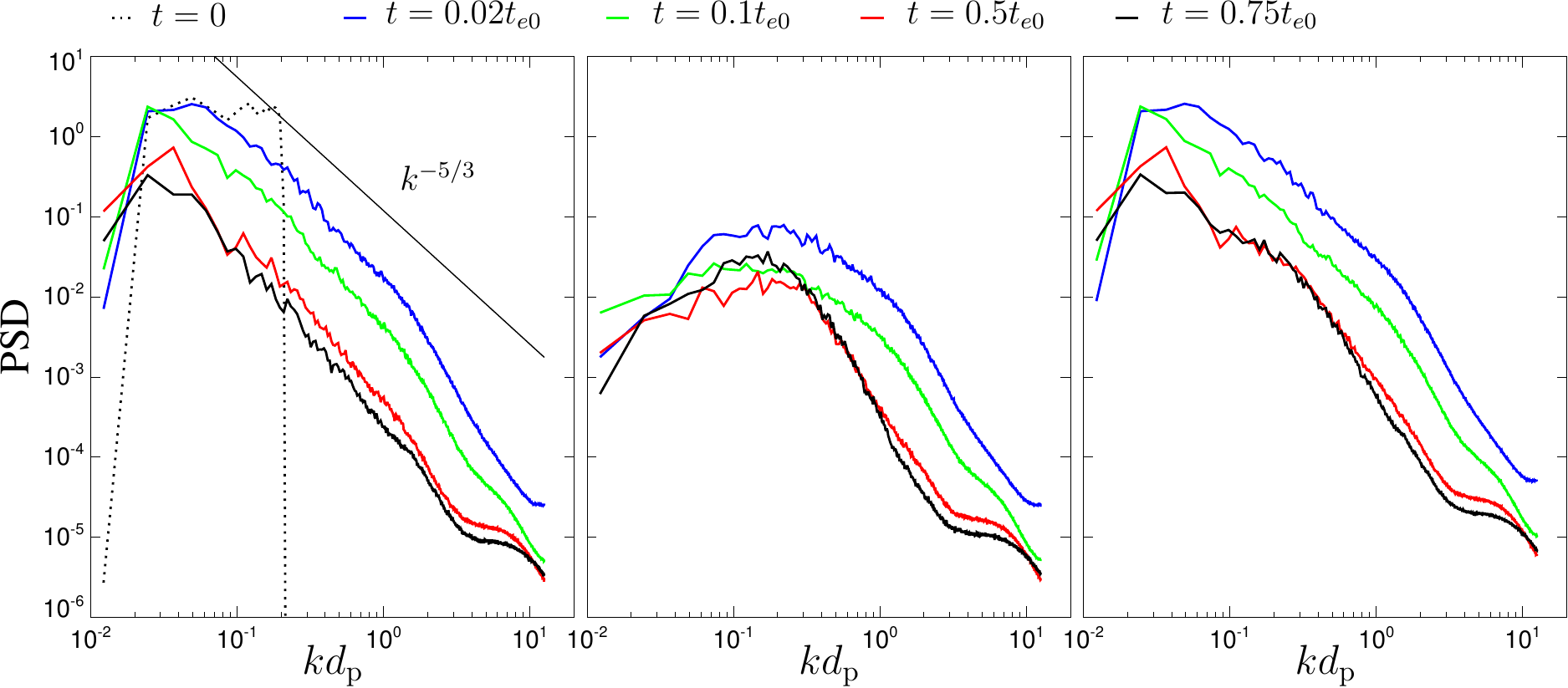}}
\caption{
1-D PSD of the  (left) perpendicular
 $B_\perp$, (middle) parallel  $B_\|$, and total (right) $B$
fluctuating magnetic field,
 normalized to $B_0^2 d_{\mathrm{p}0}$,
 as functions of $k$ at different times.
The dotted line shows the initial spectrum and
the thin solid line
shows a dependence $\propto k^{-5/3}$ for comparison.
\label{spec}}
\end{figure*}

Figure~\ref{spec} presents 1-D power spectral density (PSD) of the 
(left) perpendicular $B_\perp$, (middle) parallel $B_\|$, and total (right) $B$
 as functions of $k$ at $t=0$ (dotted line), $t=0.02 t_{e0}$ (blue)
 $t=0.1 t_{e0}$ (green), $t=0.5 t_{e0}$ (red), and $t=0.75 t_{e0}$ (black solid).
The perpendicular component $B_\perp$ has roughly a Kolmogorov-like slope
on large scales that steepens on the sub-ion scales.
The slopes of $B_\perp$, $B_\|$, and $B$ in the sub-ion range (below the transition/break) are quite
similar, about $-3.5$. However, the range where the spectra are power-law like is quite narrow,
at smaller scales $k\gtrsim 4/d_{\mathrm{p}}$ they
flatten (especially at later times) indicating a bottleneck problem possibly connected with the numerical
noise. 
The amplitude of the $B_\perp$ spectra decreases with time owing to the cascade
and the expansion. 
The amplitude of the compressible ($B_\|$) spectrum also initially decreases
but, at later times, the level of fluctuations on large scales increases
(see Figure~\ref{evol}a).
The incompressible  $B_\perp$ fluctuations dominate the total
spectrum but, at later times, the compressible component $B_\|$
becomes important on relatively large scales and importantly contributes
to the total power spectra of $B$.

\begin{figure*}[htb]
\centerline{\includegraphics[width=14cm]{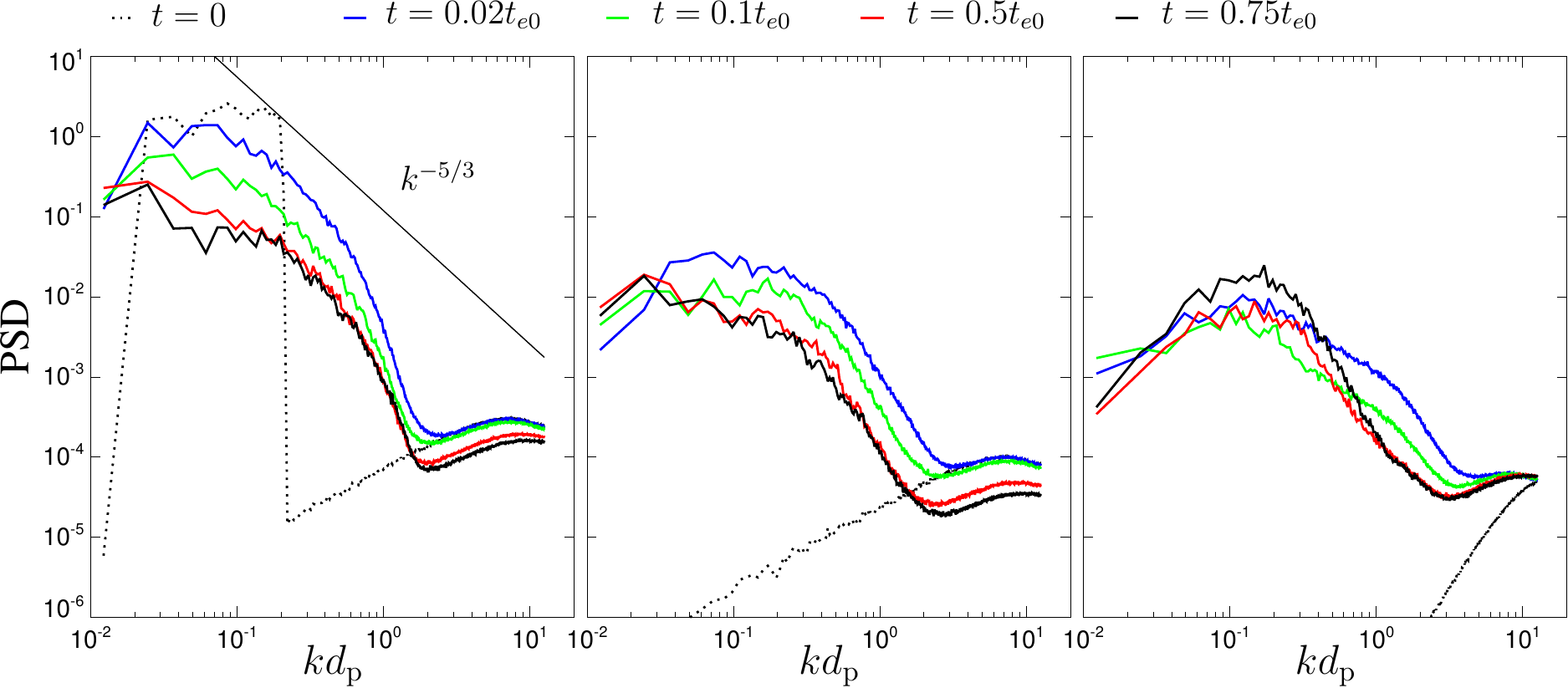}}
\caption{
1-D PSD of the fluctuating proton velocity field
 $u_\perp$ (left), $u_\|$ (middle), normalized to $v_{A0}^2 d_{\mathrm{p}0}$, and of the
density fluctuations $\delta n$ (right), normalized to $\langle n\rangle^2 d_{\mathrm{p}0}$,
 as functions of $k$ at different times.
The dotted line shows the initial spectra and
the thin solid line
shows a dependence $\propto k^{-5/3}$ for comparison.
\label{specun}}
\end{figure*}

Figure~\ref{specun} presents 1-D PSD of the proton velocity field
 $u_\perp$ (left) and $u_\|$ (middle), and the proton (number) density
(right) as functions of $k$ at $t=0$ (dotted line), $t=0.02 t_{e0}$ (blue)
 $t=0.1 t_{e0}$ (green), $t=0.5 t_{e0}$ (red), and $t=0.75 t_{e0}$ (black solid).
The power spectra of $u_\perp$ exhibit an exponential-like behavior from large
to sub-ion scales (alternatively, this may be a smooth transition between
two power-law like dependencies but it's hard to distinguish between the two
for the relatively short range of wave vectors); for $k d_{\mathrm{p}}\gtrsim 2$ the spectra are dominated
by the numerical noise due to the finite number of particles per cell \citep{franal15b}.
The power spectra of $u_\|$ are relatively flat at large scales and below
about $k \rho_\mathrm{p}$ they exhibit rather power-law like properties (with a slope about $-3$) and again,
for $k d_{\mathrm{p}}\gtrsim 2$ the spectra are dominated
by the numerical noise.
The amplitudes of $u_\perp$ and $u_\|$ decreases with time but at later times they are roughly
constant.
The power spectra of $\delta n$ at early times have properties of two power laws (but
the ranges of wave vectors are too short to be sure) with a relatively thin transition,
and their amplitude decreases in the time.
At later times the amplitude of density fluctuations increases and has a property of
a wide spectral peak with the maximum around $k d_{\mathrm{p}} \sim 10^{-1}$.
As in the case of the velocity fluctuations the density spectra are dominated
by the numerical noise for $k d_{\mathrm{p}}\gtrsim 2$.

\begin{figure*}[htb]
\centerline{\includegraphics[width=14cm]{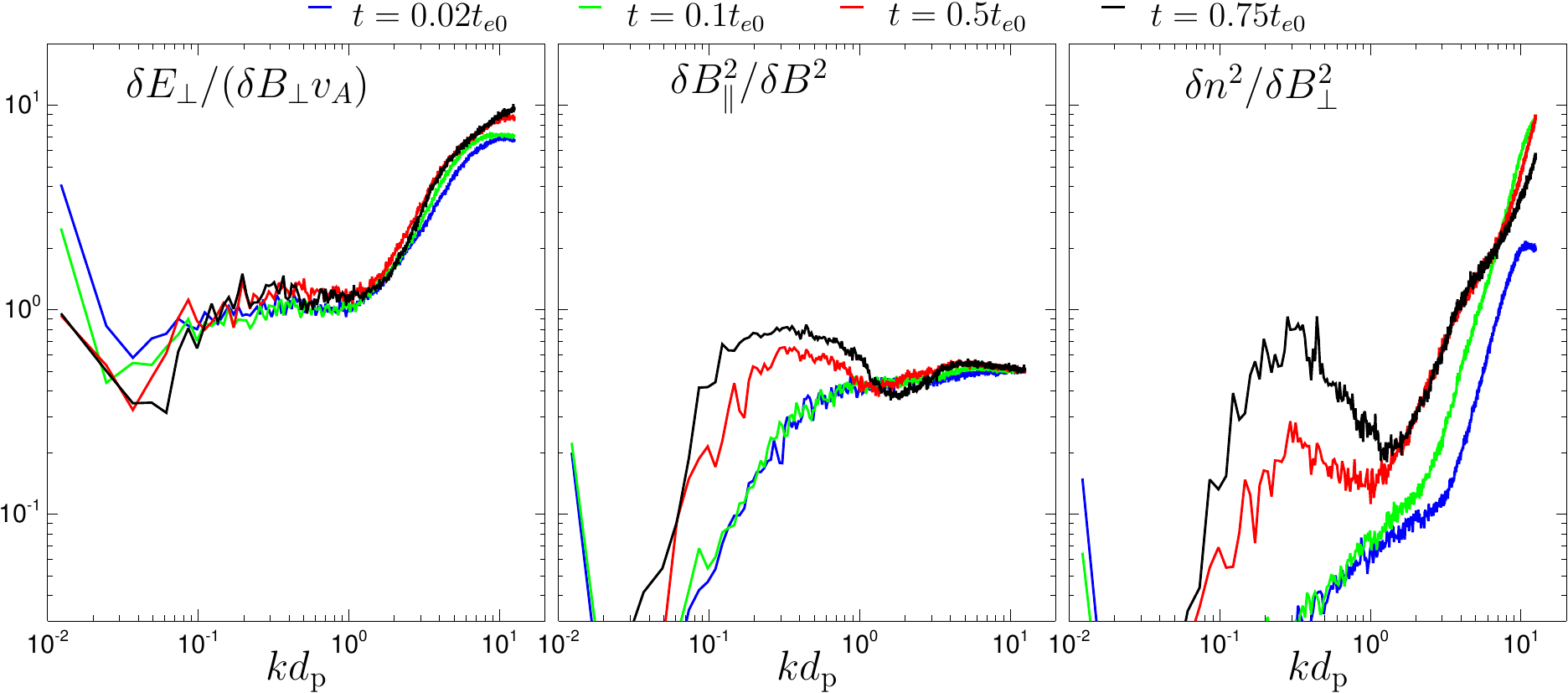}}
\caption{(left) The ratio between perpendicular electric and magnetic fluctuations,
(middle) 
the ratio between squared amplitudes of the parallel 
and the total magnetic fluctuations,
and (right) 
the ratio between squared amplitudes of density and perpendicular magnetic
fluctuations
as functions of $k$ at different times.
\label{rats}}
\end{figure*}

The generation of compressible fluctuations affects some
ratios used to analyze properties of turbulent fluctuations.
Figure~\ref{rats} shows (left) the ratio between perpendicular electric and magnetic fluctuations,
(middle)  the ratio between squared amplitudes of the parallel
and the total magnetic fluctuations, and (right)
the ratio between squared amplitudes of density and perpendicular magnetic
fluctuations as functions of $k$ at different times.
The simulation results show that the transverse fluctuations are not strongly
affected by the presence of the compressible fluctuations \cite[cf.,][]{baleal05,mattal17}
 whereas, unsurprisingly,
the compressible ratios $\delta n^2/\delta B_\perp^2$ and $\delta B_\|^2/\delta B^2$
are strongly affected \cite[cf.,][]{kiyaal13,franal15a}. 

\begin{figure}[htb]
\centerline{\includegraphics[width=8cm]{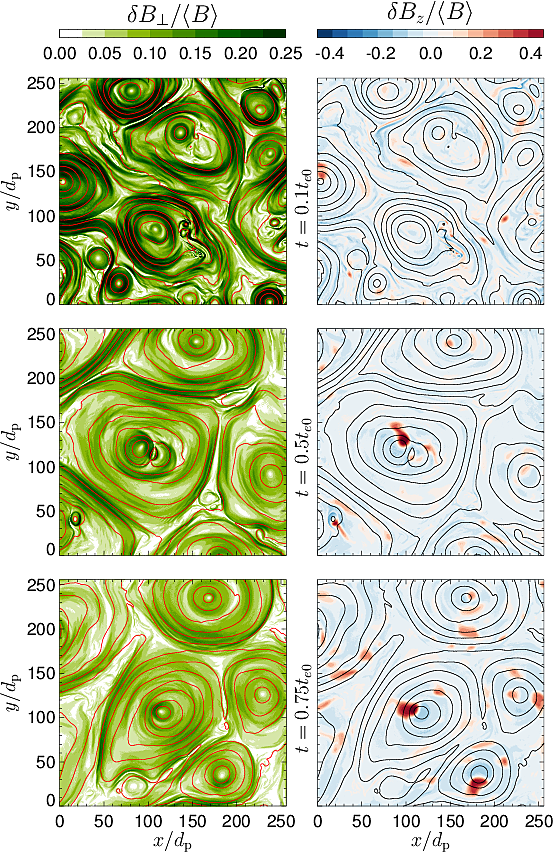}}
\caption{Color scale plots of (left) $\delta B_\perp$ and (right) $\delta B_z$
 as functions of $x$ and $y$ for (top) $t=0.1 t_{e0}$,
(middle) $t=0.5 t_{e0}$,  and (bottom) $t=0.75 t_{e0}$. The solid lines
show selected (projected) magnetic field lines.
 Only a part of the simulation box is shown.
\label{bb}}
\end{figure}

The spatial properties of the magnetic fluctuations are shown in Figure~\ref{bb}
where $\delta B_\perp$ and $\delta B_z$ are displayed as functions of $x$ and $y$
at different times: (top) $t=0.1 t_{e0}$,
(middle) $t=0.5 t_{e0}$,  and (bottom) $t=0.75 t_{e0}$.
Only a part of the simulation box is shown; note that the radial, $y$ size of
the simulation box normalized to $d_\mathrm{p}$ decreases in time as $d_\mathrm{p}\propto
R$ (see the animation corresponding to Figure~\ref{bb}). 
The slow expansion introduces an anisotropy with respect to the radial direction
 \citep{dongal14,vegr15,vegr16}; this happens mainly on the large scales,
the turbulent characteristic time scales on the scales resolved in the present simulation are much faster
then the expansion time so that no clear anisotropy is observed in our simulation \cite[cf.,][]{vech16}. 
Figure~\ref{bb} shows the turbulent field of magnetic islands/vortices in $\delta B_\perp$ and
formation of localized magnetic enhancements/humps
in the compressible magnetic component  $\delta B_z$ that are evident at later times
but weak signatures of these structures are already seen at $t=0.1 t_{e0}$. The compressible structures
 are likely the expected mirror mode structures, a more detailed analysis indicates that
these structures are standing in the local plasma frame (they are moving with the turbulent
plasma flow, see the animation corresponding to Figure~\ref{bb}).

\begin{figure}[htb]
\centerline{\includegraphics[width=8cm]{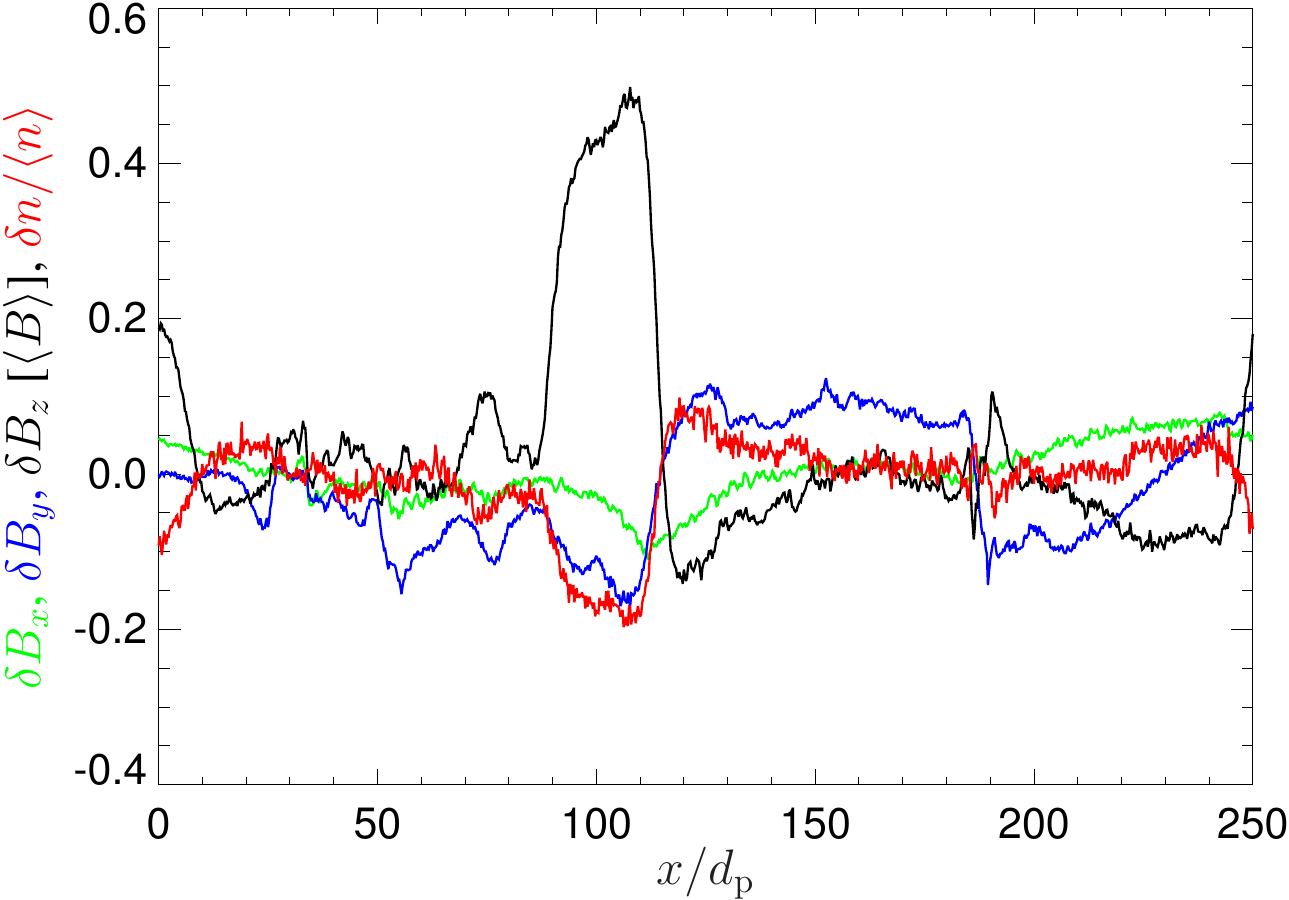}}
\caption{Spatial profiles of $\delta B_x$,  $\delta B_y$, $\delta B_z$ 
(all normalized to $\langle B \rangle$), and $\delta n$ (normalized to $\langle n \rangle$)
 as functions of $x$ at $y=110 d_{\mathrm{p}}$ for $t=0.75 t_{e0}$. 
 (see Figure~\ref{bb}, bottom).
\label{profb}}
\end{figure}

A detailed view of the spatial structure is displayed in Figure~\ref{profb}
showing 1-D cuts of $\delta B_x$,  $\delta B_y$, $\delta B_z$ 
(all normalized to $\langle B \rangle$), and $\delta n$ (normalized to $\langle n \rangle$) 
as functions of $x$ at $y=110 d_{\mathrm{p}}$ and $t=0.75 t_{e0}$ (see Figure~\ref{bb}, bottom).
 $\delta B_x$ and  $\delta B_y$ components of the fluctuating magnetic field have a
complex structure, at around $x=110 d_\mathrm{p}$ the cut passes a center of a relatively large
magnetic vortex. Close to this center the compressible component $\delta B_z$ forms a
magnetic hump with a strong amplitude $\delta B_z/\langle B \rangle \sim 0.5$. The magnetic
enhancement is compensated by a density decrease; the magnetic hump/density hole structure is roughly at 
pressure balance.

\begin{figure}[htb]
\centerline{\includegraphics[width=8cm]{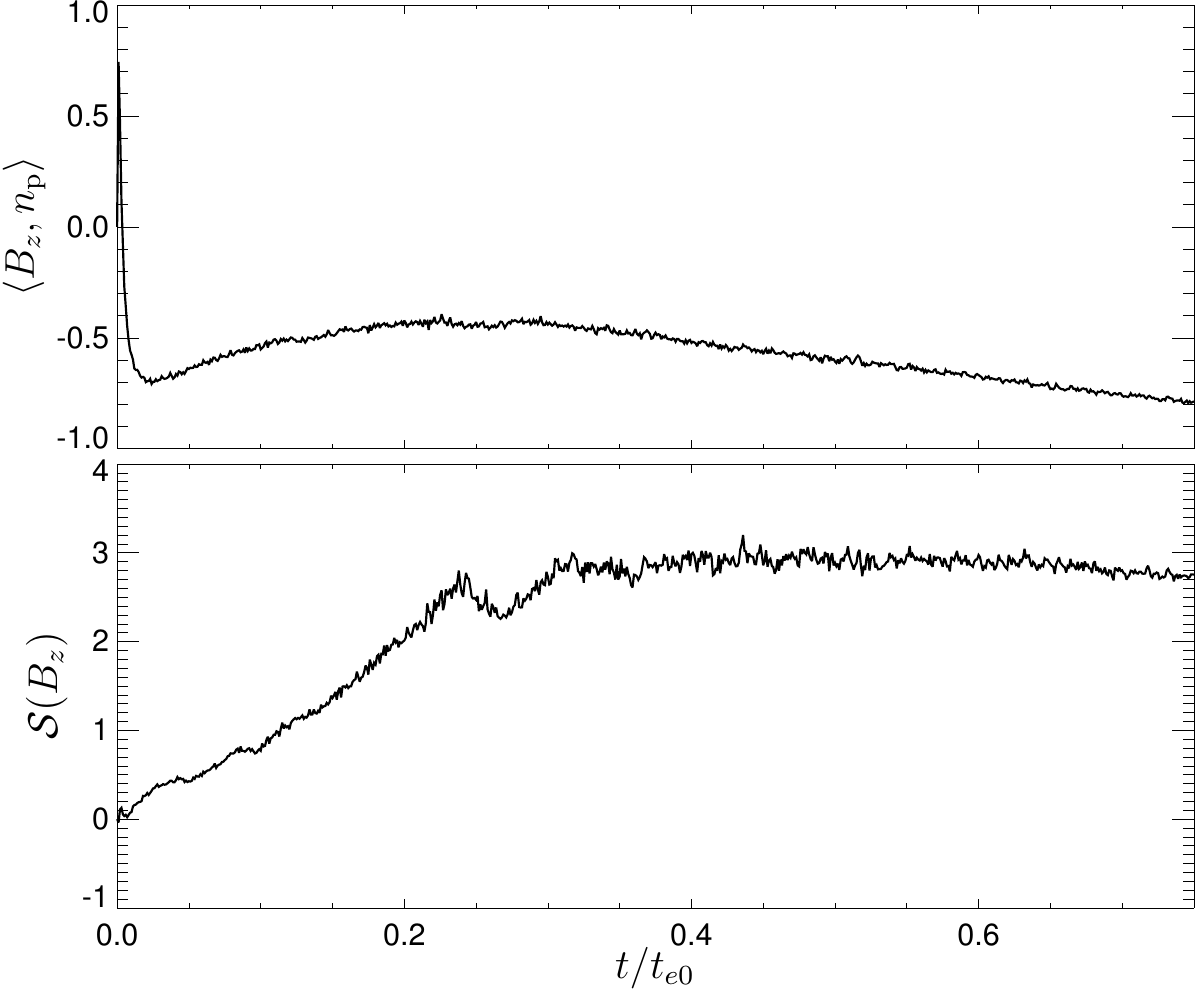}}
\caption{Time evolution of different quantities: (top) the
correlation between the compressible component $B_z$ and the proton number density $n_\mathrm{p}$ and (bottom)
of the skewness of $B_z$, $\mathcal{S}(B_z)$, as functions of time.
\label{corskew}}
\end{figure}

The pressure-balanced magnetic humps are characterized by an anti-correlation between
the magnetic field component $B_z$ and the proton density and the distribution of $B_z$
values has a skewed distribution with a positive skewness \citep{genoal09}.
Figure~\ref{corskew} shows the evolution of the correlation between $n$ and $B_z$
and the skewness of $B_z$, $\mathcal{S}(B_z)$, calculated over the whole box, as functions of time.  
Initially in the simulation, $B_z$ and $n$ are correlated indicating fast mode-like properties.
 $B_z$ and $n$ rapidly become anti-correlated suggesting a slow mode-like behavior; a similar
evolution is seen in standard 2-D hybrid simulations of \cite{franal16b}. At later times
the anti-correlation is strengthened by development of mirror structures.
The skewness $\mathcal{S}(B_z)$ starts around zero, then steadily increases, and for
$t\gtrsim 0.3  t_{e0}$ saturates around $3$. It is interesting to note that 
a similar analysis applied to the results from standard 2-D hybrid simulations of
\cite{franal16b} shows that the skewness  
 $\mathcal{S}(B_z)$ is negative in a
well-developed turbulent cascade starting from an isotropic proton distribution 
for a wide range of betas.

\begin{figure}[htb]
\centerline{\includegraphics[width=8cm]{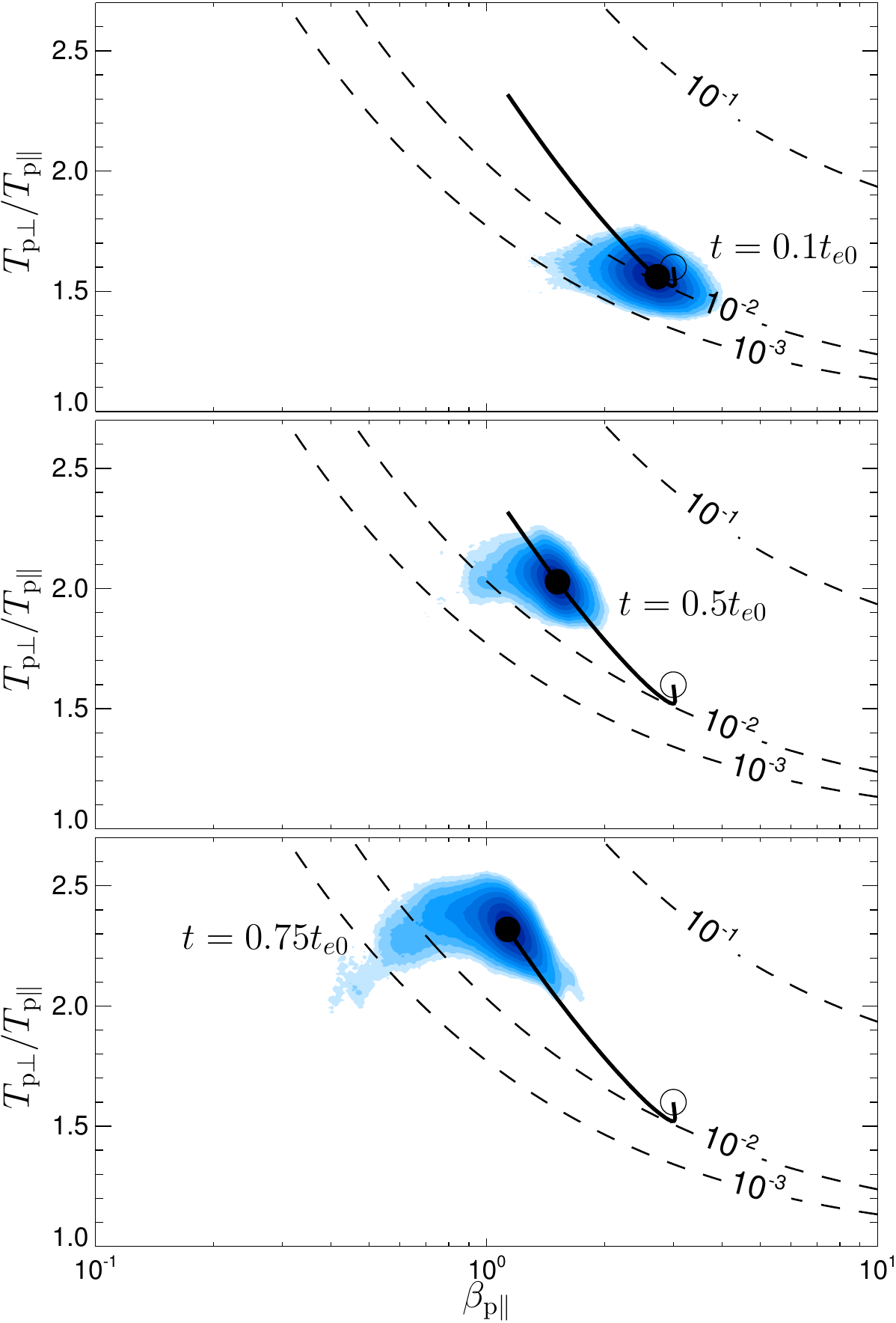}}
\caption{Simulated data distribution in the plane
$(\beta_{\mathrm{p}\|},T_{\mathrm{p}\perp}/T_{\mathrm{p}\|})$ at different times.
The empty circles give the initial condition whereas the solid circles
denote the average values and the solid
lines show their evolution.
The dashed contours show
 the maximum growth rate $\gamma_{max}$ (in units of $\Omega_\mathrm{p}$) of the mirror 
instability
as a function of $\beta_{\mathrm{p}\|}$ and $T_{\mathrm{p}\perp}/T_{\mathrm{p}\|}$
in the corresponding plasma with bi-Maxwellian protons.
\label{bean}}
\end{figure}

\begin{figure}[htb]
\centerline{\includegraphics[width=8cm]{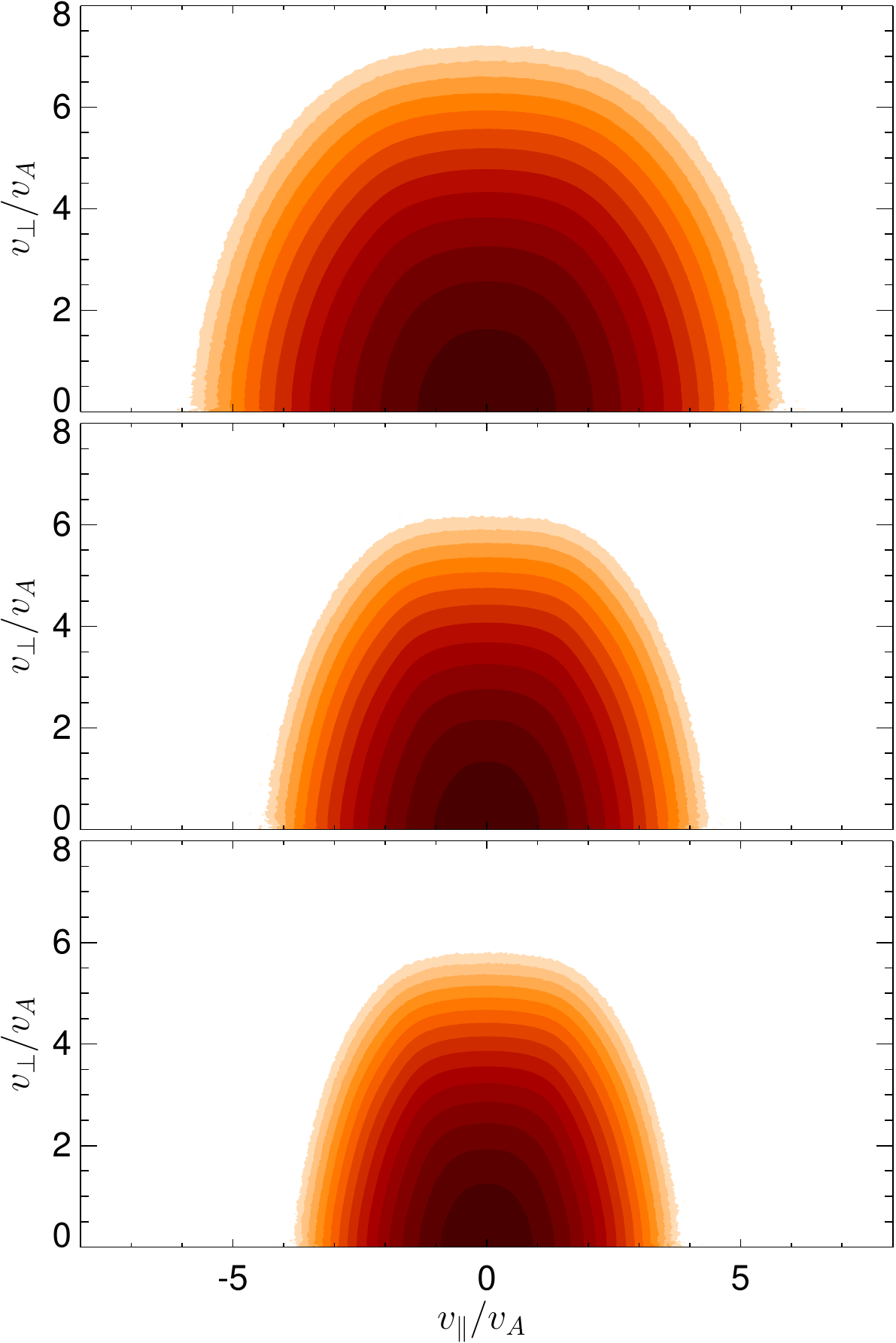}}
\caption{Average proton velocity distribution function $f$
as a function of parallel and perpendicular velocities $v_\|$ and $v_\perp$
(with respect to the local magnetic field) for (top) $t=0.1 t_{e0}$,
(middle) $t=0.5 t_{e0}$,  and (bottom) $t=0.75 t_{e0}$.
\label{vdf}}
\end{figure}

Figure~\ref{bean} shows the simulation results
$(\beta_{\mathrm{p}\|},T_{\mathrm{p}\perp}/T_{\mathrm{p}\|})$ at different times,
(top) $t=0.1 t_{e0}$, (middle)  $t=0.5 t_{e0}$, and (bottom) $t=0.75 t_{e0}$.
The shades of blue display the distribution of the local (grid) values
whereas the solid circles indicate the values averaged over the simulation box.
The empty circle denotes the initial condition and the solid line gives the
evolution of the averaged values.
During the evolution, a large spread of local values develops in the
space $(\beta_{\mathrm{p}\|},T_{\mathrm{p}\perp}/T_{\mathrm{p}\|})$ 
\cite[cf.,][]{serval15,hellal15}. The expansion drives the system towards
more unstable situations but the development of mirror modes reduces
the anisotropy and tends to stabilize the system. A small subset of local
values in the space $(\beta_{\mathrm{p}\|},T_{\mathrm{p}\perp}/T_{\mathrm{p}\|})$
have smaller values of the maximum growth rates (compared to the average value);
the places where the mirror instability is weakened appear in the vicinity
of the mirror structures. The reduction of the local proton temperature
anisotropy is mainly governed by the enhanced magnetic field that
leads to higher proton perpendicular temperatures, the magnetic moment of
protons is varying only weakly \cite[cf.,][]{scheal08}.
The variation of the magnetic field is not, however, the only saturation mechanism.
Figure~\ref{vdf} shows the proton velocity distribution $f$ (averaged over the simulation box)
as a function of parallel and perpendicular velocities $v_\|$ and $v_\perp$
(with respect to the local magnetic field)
 at different times:
(top) $t=0.1 t_{e0}$,
(middle) $t=0.5 t_{e0}$,  and (bottom) $t=0.75 t_{e0}$.
The averaged proton distribution function exhibit a clear flattening 
around $v_\|=0$, i.e., $\partial f/\partial v_\| \sim 0$ for $v_\perp \gtrsim 2 v_A$.   
This is compatible with the quasi-linear diffusion of protons through 
the Landau (transit time) resonance
\citep{calial08,hellal09}.

Both the linear and nonlinear properties of the mirror instability in the
nonlinear reductive perturbative model 
\citep{kuznal07b, kuznal07a,calial08} are sensitive to the
details of the proton distribution function in the resonant
region $v_\|\sim 0$.
 The flattening observed in the simulation
 likely modifies the nonlinear properties
and leads to generation of magnetic humps (instead of holes that
are expected for bi-Maxwellian velocity distribution functions \citep{hellal09})
in agreement with the simulation results.

\section{Discussion}
\label{discussion}

The presented 2-D hybrid simulation of plasma turbulence with 
the expansion forcing demonstrates that mirror instability  may coexist
with the fully developed (strong) turbulence and generate compressible,
nonpropagating
pressure-balanced magnetic structures with amplitudes comparable to or even
greater than those of the ambient turbulent fluctuations. 
The compressible component of the magnetic field $B_\|$ 
becomes at later times important on scales comparable to and larger
than the typical proton scales and this
affects the total power spectra of $B$ around the transition
between large MHD and sub-ion scales \cite[cf.,][]{lional16},
 as well as different compressibility ratios.
The mirror structures reduce locally the temperature anisotropy through
a fluid mechanism (generation of enhanced magnetic field that
increases the proton perpendicular temperature) and through a 
quasilinear-like proton scattering via the Landau (transit time)
resonance. The role of particle trapping in the present case is
unclear \citep{rincal15}.

The dominant, compressible magnetic component of
the mirror structures only weakly interacts with the incompressible turbulent Alfv\'enic 
fluctuations through the main fluid nonlinearities.
On the other hand, the minor, transverse component of mirror fluctuations have a vortex like
properties around the compressible structures \citep{passal14} and  likely couples directly
to the turbulent plasma motions (the present simulation indicates
 presence of such vortical structures). Further work is necessary to
understand the interaction between turbulence and the mirror instability.

In this paper we drive the temperature anisotropy by the expansion with the
transverse magnetic field. We expect a similar evolution for other driving
forces that generate the perpendicular proton/ion temperature anisotropy 
\citep{kunzal14}.
Our work is relevant mainly for high beta plasmas where
the mirror instability is dominant; for low and moderate beta plasmas the ion cyclotron
instability is prevalent \citep{gary92,labe95}. 
The nonlinear competition between these instabilities is a nontrivial problem even in a homogeneous system
and generally requires fully three-dimensional (3-D) simulations \citep{shojal09}.
In the present case, both turbulence and the 2-D geometry constraints strongly affect
the dynamics of the mirror instability. In the 2-D simulation box we have
only limited access to oblique modes; however, the mirror instability
appear at strongly oblique angles with respect to the ambient magnetic
field (in a homogeneous plasma system) near threshold \citep{hell07}.
The 2-D geometry is likely a smaller problem for the mirror instability
compared to the oblique fire hose \cite[cf.,][]{hellal15} but, in any case,
3-D simulations are needed to
investigate the interplay between turbulence and instabilities.
We expect that  3-D simulations with turbulent fluctuations will exhibit
an evolution similar to the 2-D simulation of \cite{traval07b} modified by turbulence;
this will be subject of future work.
Despite the limitations, the present simulation results confirm that the mirror instability
is a viable mechanism that can generate magnetic pressure-balanced structures
in turbulent astrophysical plasmas.

\acknowledgments
The authors wish to acknowledge valuable idea exchanges 
with T.~Tullio.
P.H.~acknowledges grant 15-10057S of the Czech Science Foundation.
L.F. is funded by Fondazione Cassa di Risparmio di Firenze, through the project ``Giovani Ricercatori Protagonisti''.
L.M. was funded by the UK STFC grant ST/N000692/1. 
The (reduced) simulation data are available at the
Virtual Mission Laboratory Portal (http://vilma.asu.cas.cz)
developed within the
European Commission's 7\textsuperscript{th} Framework Programme under
the grant agreement \#284515 (project-shock.eu).

\end{document}